\documentclass[letterpaper, 10 pt, conference]{ieeeconf}
\IEEEoverridecommandlockouts
\usepackage{amsmath}
\usepackage{amssymb}
\usepackage{graphicx}

\usepackage[utf8]{inputenc}
\usepackage[T1]{fontenc}

\title{\LARGE \bf
Cross-Modality Translation with Generative Adversarial Networks to Unveil Alzheimer's Disease Biomarkers
}

\author{
    Reihaneh Hassanzadeh$^{1,2}$, Anees Abrol$^{2}$, Hamid Reza Hassanzadeh$^{3}$, and Vince D. Calhoun$^{1,2}$ 
    \thanks{$^{1}$School of Electrical and Computer Engineering, Georgia Institute of Technology, Atlanta, GA 30302, USA} 
    \thanks{$^{2}$Tri-institutional Center for Translational Research in Neuroimaging and Data Science (TReNDS), Georgia State University, Atlanta, GA 30303, USA}
    \thanks{$^{3}$Courtesy Faculty Appointment, College of Pharmacy, University of Florida}
}

\begin{document}

\maketitle
\thispagestyle{empty}
\pagestyle{empty}

\begin{abstract}
Generative approaches for cross-modality transformation have recently gained significant attention in neuroimaging. While most previous work has focused on case-control data, the application of generative models to disorder-specific datasets and their ability to preserve diagnostic patterns remain relatively unexplored. Hence, in this study, we investigated the use of a generative adversarial network (GAN) in the context of Alzheimer's disease (AD) to generate functional network connectivity (FNC) and T1-weighted structural magnetic resonance imaging data from each other. We employed a cycle-GAN to synthesize data in an unpaired data transition and enhanced the transition by integrating weak supervision in cases where paired data were available. Our findings revealed that our model could offer remarkable capability, achieving a structural similarity index measure (SSIM) of $0.89 \pm 0.003$ for T1s and a correlation of $0.71 \pm 0.004$ for FNCs. Moreover, our qualitative analysis revealed similar patterns between generated and actual data when comparing AD to cognitively normal (CN) individuals. In particular, we observed significantly increased functional connectivity in cerebellar-sensory motor and cerebellar-visual networks and reduced connectivity in cerebellar-subcortical, auditory-sensory motor, sensory motor-visual, and cerebellar-cognitive control networks. Additionally, the T1 images generated by our model showed a similar pattern of atrophy in the hippocampal and other temporal regions of Alzheimer's patients.

\end{abstract}

\section{Introduction}
Brain imaging techniques provide a multifaceted view of the brain. For example, structural magnetic resonance imaging (sMRI) illuminates the structure of the brain \cite{vlaardingerbroek2013magnetic}, functional MRI (fMRI) provides a dynamic picture of brain activity \cite{logothetis2008we}, and positron emission tomography (PET) delves into the metabolic processes of the brain \cite{phelps2000positron}. Although each of these modalities contributes profound insights, their integration through multimodal analysis provides a more comprehensive understanding of the brain \cite{calhoun2016multimodal}.

There is, however, a major limitation in conducting a multimodal study on real datasets: the availability of different modalities at varying scales. That is, available multimodal datasets often exhibit missing modalities for many subjects, which can be attributed to a variety of factors such as ease of data collection and implementation costs associated with each modality. For example, fMRI requires a subject to remain still for a relatively longer duration than sMRI and hence be more challenging for infants or individuals with conditions like Alzheimer’s disease. Thus, existing neuroimaging datasets often contain fewer samples of particular modalities, requiring efficient approaches to handle missing data in multimodal analysis.

Conventionally, approaches like mean imputation, downsampling, or zero-filling \cite{venugopalan2021multimodal} have been employed to facilitate multimodal analyses when faced with incomplete modalities in datasets. While these techniques are relatively simple to apply, their effectiveness is compromised due to either the omission of essential data or the inadvertent creation of bias by altering the genuine distribution of the data. This calls for more advanced approaches that not only leverage the available data but also ensure a more accurate analysis. 

Recently, advanced techniques, such as generative modeling, have emerged as promising alternatives to complete missing data while not having such limitations. Previous work developed GAN-based models for modality translation in PET-sMRI \cite{cai2018deep,gao2021task}, EEG-fMRI \cite{cheng2021research}, multi-contrast sMRI \cite{tiago2023domain,dar2019image,yuan2020unified}, and sMRI-CT \cite{tiago2023domain,yuan2020unified}. While most of the existing works on cross-modality transformation have examined a controlled-case dataset, the application of generative models to disorder-specific datasets and their ability to capture diagnostic patterns is more challenging and an area that remains relatively unexplored. 

\begin{figure*}[thpb]
    \begin{center}
        \includegraphics[width=0.7\textwidth]{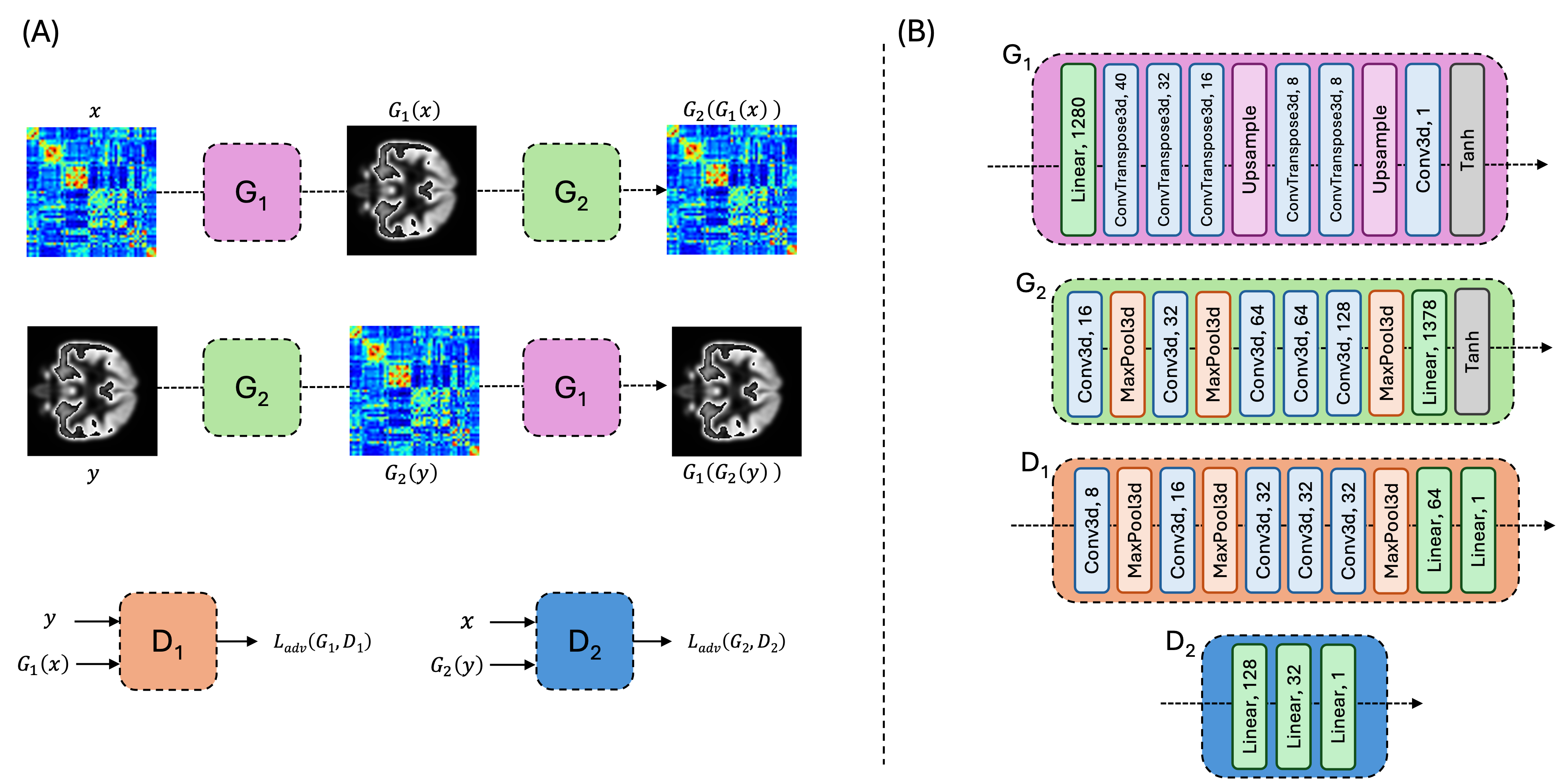}
        \centering
        \caption{(A) Translation framework. (B) Architecture of generators and discriminators. Numbers indicate the output channel and linear size.}
        \label{framework}
    \end{center}
\end{figure*}

In this study, we explored the ability of generative models for T1-FNC translation within the context of Alzheimer's disease. We employed cycle-GAN to synthesize FNC and T1 from each other in an unpaired data transition. To enhance the efficacy of our model, we integrated weak supervision \cite{zhu2017unpaired} during the transition when paired data were available (i.e., both modalities were present). We evaluated our model both quantitatively using established metrics such as the SSIM \cite{wang2004image} and Pearson correlation that measure the accuracy of the translations and qualitatively through visual assessments of the generated images against the original.  Our findings showed that our model can successfully transform the tested neuroimaging modalities while preserving functional and structural brain patterns specific to Alzheimer's disease.

\section{Method}
\subsection{Cross-Modality Transformation Framework}
We utilized Cycle-GAN \cite{zhu2017unpaired}, which uses unpaired data, and its objective is to translate data from a domain X (e.g., FNC maps) into another domain Y (e.g., T1 images) and vice versa. Our approach, shown in Fig. \ref{framework}.A, is an extension to that model where we use two generators $G_1$ and $G_2$ that map one modality into another and two discriminators $D_1$ and $D_2$ that differentiate the actual data from the generated ones. Each of the generators and its corresponding discriminator were trained in an adversarial manner. In other words, each generator tries to construct samples that resemble the real data, and its discriminator tries to identify whether a sample is produced by the generator. We used the least square adversarial loss to train our model as formulated below:
\begin{equation}
    \mathcal{L}_{adv}(G_1, D_1) = 
    \begin{aligned}[t]
        &\mathbb{E}_{y \sim p_{\text{data}}(y)} \left[\left(D_1(y) - 1\right)^2\right] \\
        &+ \mathbb{E}_{x \sim p_{\text{data}}(x)} \left[\left(D_1(G_1(x))\right)^2\right],
    \end{aligned}
\end{equation}
\begin{equation}
    \mathcal{L}_{adv}(G_2, D_2) = 
    \begin{aligned}[t]
        &\mathbb{E}_{x \sim p_{\text{data}}(x)} \left[\left(D_2(x) - 1\right)^2\right] \\
        &+ \mathbb{E}_{y \sim p_{\text{data}}(y)} \left[\left(D_2(G_2(y))\right)^2\right].
    \end{aligned}
\end{equation}

Optimizing the adversarial loss alone may lead to mode collapse \cite{shrivastava2017learning}. To partially mitigate this issue, the cycle consistency loss has been employed in literature to regularize generators' mapping process and enforce cycle consistency. The cycle-consistency loss is formulated as follows:
\begin{equation}
    \mathcal{L}_{cyc}(G_1, G_2) = 
    \begin{aligned}[t]
        &\mathbb{E}_{x \sim p_{\text{data}}(x)} \left[\left\| G_2(G_1(x)) - x \right\|_1\right] \\
        &+ \mathbb{E}_{y \sim p_{\text{data}}(y)} \left[\left\| G_1(G_2(y)) - y \right\|_1\right].
    \end{aligned}
\end{equation}

Additionally, we incorporated weak supervision whenever paired data were available to achieve a more accurate translation. This supervision was implemented using an identity loss as stated below:
\begin{equation}
    \mathcal{L}_{identity}(G_1, G_2) = 
    \begin{aligned}[t]
        &\sum_{(x,y) \in \mathcal{P}} \left[\left\| G_1(x) - y \right\|_1\right] \\
        &+\sum_{(x,y) \in \mathcal{P}} \left[\left\| G_2(y) - x \right\|_1\right],
    \end{aligned}
\end{equation}

where $\mathcal{P}$ is the set of paired data.




Overall, the final objective loss becomes:
\begin{equation}
    \mathcal{L}(G_1, G_2, D_1, D_2) = 
    \begin{aligned}[t]
        &\mathcal{L}_{adv}(G_1, D_1) \\
        &+ \mathcal{L}_{adv}(G_2, D_2) \\
        &+ \lambda_1 \mathcal{L}_{cyc}(G_1, G_2) \\
        &+ \lambda_2 \mathcal{L}_{identity}(G_1, G_2),
    \end{aligned}
\end{equation}
where $\lambda_1$ and $\lambda_2$ are trade-off hyperparameters controlling the effect of their corresponding loss in the optimization process.

\begin{figure*}[thpb]
    \begin{center}
        \includegraphics[width=0.75\textwidth]{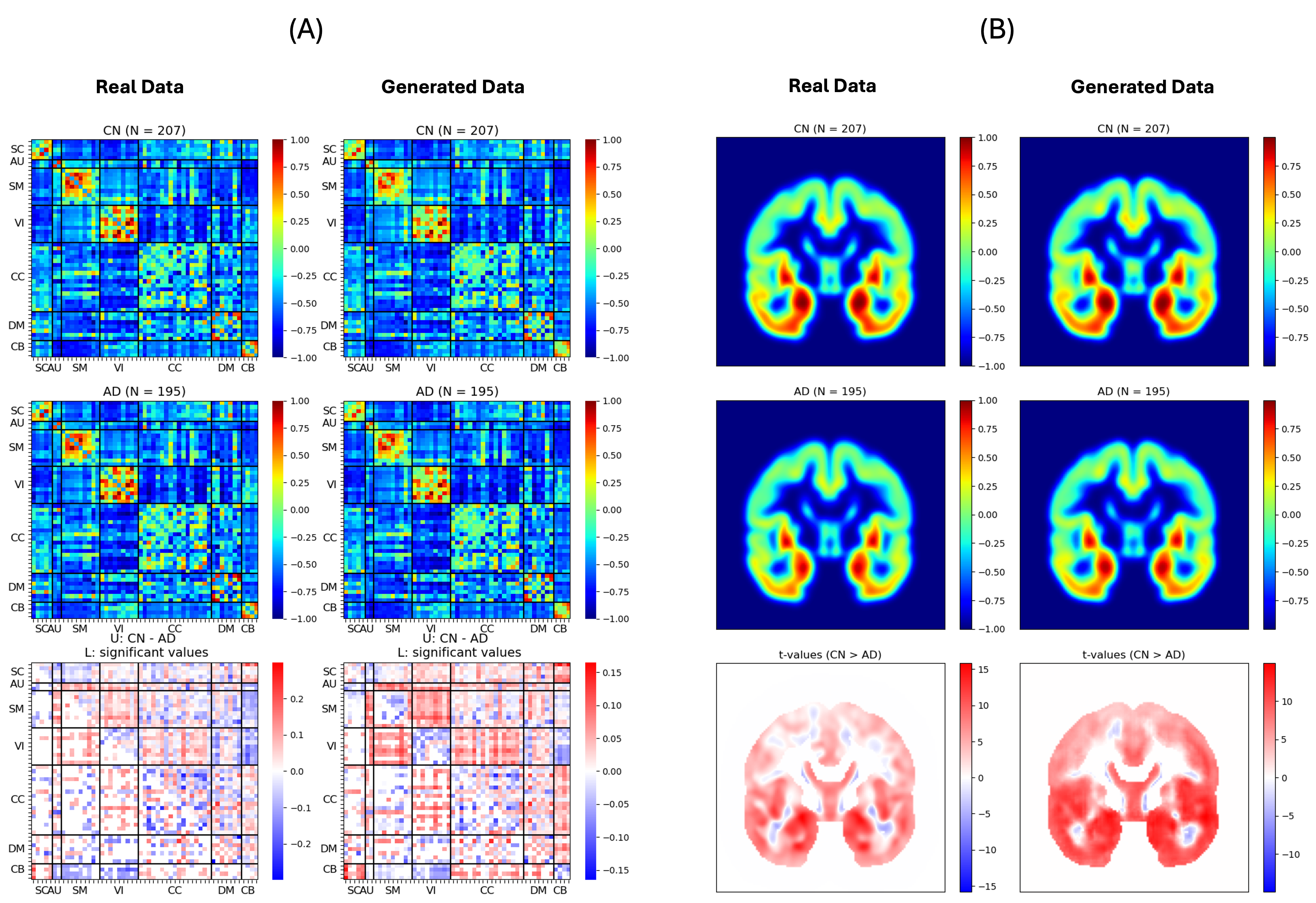}
        \centering
        \caption{(A) Real (first column) and generated (second column) FNC maps and the group mean differences. (B) Real and generated T1 images and $t$-values. Numbers in parentheses represent the sample size, and U and L indicate the upper and lower triangular matrix, respectively. Note that, to compare real and generated samples, we visualized only the samples for which both FNC and T1 data were available.}
        \label{group-diff}
    \end{center}
\end{figure*}

\subsection{Model Architecture} 
Fig. \ref{framework}.B provides an overview of the model architectures employed for the generators and discriminators in our model. The generator $G_1$ is designed to transform 1D-FNC maps, each of size 1378, into 3D-T1 images of dimensions 121x145x121 through a sequence of five 3D-transposed convolution layers, each followed by batch normalization, and a final convolutional layer. The outputs of the last layer were then scaled with the tanh activation. We added two upsampling layers to match the final output size to the real T1 images. $G_2$ maps T1 images into FNC maps through five 3D convolutional layers followed by batch normalization and max-pooling, as illustrated in the figure. The outputs of the last convolutional network were transformed into FNC maps after passing a linear layer and a tanh activation. $D_1$ and $D_2$ combine several convolutional and linear layers, as illustrated in the figure.

\section{Datasets and Experiments}
We used the Alzheimer’s Disease Neuroimaging Initiative (ADNI) dataset \cite{jack2008alzheimer}, including a total of 2923 images from 986 subjects. Table \ref{table:data-dist} shows the distribution of samples and subjects across modalities and diagnosis groups. While the dataset includes 2910 T1 images, only 414 FNC maps are available. 

\begin{table}[h!]
\caption{Data Distribution.}
\centering
\begin{tabular}{|c|c|c|c|}
\hline
Diagnosis & \multicolumn{3}{c|}{\# of Images (\# of Subjects)} \\ \cline{2-4} 
                           & FNC         & T1         & FNC and T1         \\ \hline
CN                         & 207 (122)   & 1446 (414)  & 207 (122)           \\ \hline
AD                         & 207 (94)    & 1465 (575)  & 195 (85)            \\ \hline
\end{tabular}
\label{table:data-dist}
\end{table}


The T1 data were segmented into tissue probability maps for gray matter, white matter, and cerebral spinal fluid using SPM12. The gray matter images were then warped to standard space, modulated and smoothed using a Gaussian kernel with an FWHM = 10mm. The preprocessed gray matter volume images had a dimensionality of 121×145×121 in the voxel space, with the voxel size of 1.5×1.5×1.5 $mm^3$.

For the resting-state fMRI data sets, we removed the first five time points to ensure signal equilibrium and adaptation of subjects to scanner noise. Then we performed slice timing correction and rigid body motion correction using the SPM toolbox, followed by warping the images into the standard Montreal Neurological Institute (MNI) template using an echo-planar imaging (EPI) template and the old SPM12 normalization module. Lastly, the data were resampled to 3x3x3 $mm^3$ isotropic voxels, resulting in image dimensionality of 53x63x52 in the voxel space, and smoothed using a Gaussian kernel with a full width at half maximum of 6mm. We also implemented thorough quality control (QC) on the preprocessed fMRI images to discard the images that exhibited (1) poor correlation with individual and group data masks, (2) markedly briefer scan lengths, and (3) high head motion parameters (>3° rotations and >3mm translations).

To generate FNC maps, the resting-state fMRI data were pre-processed using the NeuroMark, a fully automated independent component analysis (ICA) pipeline \cite{du2020neuromark}, resulting in 53 components. These components are grouped into seven network domains, namely subcortical (SC), auditory (AU), visual (VI), sensorimotor (SM), cognitive control (CC), default mode (DM), and cerebellar (CB). FNC maps were computed using the Pearson correlation coefficient between two ICA-estimated components.

We trained our model in a 5-fold cross-validation framework for 300 epochs. We used a batch size of 32 and an Adam optimizer with an initial learning rate of 0.05, which decayed by 0.9 every five epochs. We used $\lambda_1=10$ and $\lambda_2=40$, which were selected from a grid search within the value range of 1, 5, 10, 20, and 40. We kept the latest 50 generated samples in a buffer and updated our discriminators using the images in the buffer to reduce the potential for mode collapse further \cite{goodfellow2020generative}. We measured SSIM between generated and actual T1 images and Pearson correlation between generated and real FNC samples to quantify the accuracy of generated data from our GAN-based model. Furthermore, we evaluated the generated samples through qualitative assessments of the generated images against the original and conducted $t$-test to statistically compare the differences in mean between diagnosis groups.

\section{Results}

Table \ref{table:quantitative-metrics} shows the quantitative evaluation between the generated and the actual data. We achieved an SSIM of $0.894 \pm 0.003$ for T1s and a correlation of $0.707 \pm 0.004$ for FNCs. Comparing these metrics across diagnostic groups, we observed that our model could accurately reproduce data for both AD and CN individuals. 

\begin{table}[h]
\caption{Quantitative Evaluation Metrics}
\label{table:quantitative-metrics}
    \begin{center}
        \begin{tabular}{|c|c|c|}
            \hline
            Diagnosis & Pearson-Correlation & SSIM\\
            \hline
            All Data & 0.707$\pm$ 0.004 & 0.894$\pm$ 0.003 \\
            CN & 0.707$\pm$ 0.006 & 0.897$\pm$ 0.003 \\
            AD & 0.705$\pm$ 0.01 & 0.892$\pm$ 0.005 \\
            \hline
        \end{tabular}
    \end{center}
\end{table}

Table \ref{table:method_comparison} shows that the comparison between the proposed method and the standard cycle-GAN as the baseline, where the identity loss ($\lambda_2=0$) is excluded. Based on the results, our method can statistically significantly outperform both SSIM ($p$=0.003) and Pearson correlation ($p$=0.0008). 

Moreover, we evaluated our results qualitatively through visual assessments of the generated images against the original and observed similar AD-specific patterns between the generated and real data. In particular, according to Figure \ref{group-diff}.A, in Alzheimer’s patients compared to CN individuals, we observed significant increases in functional connectivity of the CB-SM \cite{penalba2023increased} and CB-VI network pairs and significant decreases in the CB-SC, AU-SM, SM-VI, CB-CC, and DM-DM \cite{badhwar2017resting} network pairs. Additionally, as illustrated in Figure \ref{group-diff}.B, the T1 images generated by our model revealed neural loss \cite{mckhann2011diagnosis} in the hippocampal and other temporal regions of Alzheimer’s patients with a similar atrophy pattern to the real data.

\begin{table}[h]
\caption{Method Comparison}
\label{table:method_comparison}
    \begin{center}
        \begin{tabular}{|c|c|c|}
            \hline
            Diagnosis & Pearson-Correlation & SSIM\\
            \hline
            Proposed Method & \textbf{0.707}$\pm$ 0.004 & \textbf{0.894}$\pm$ 0.003 \\
            Baseline & 0.69$\pm$ 0.006 & 0.871$\pm$ 0.012 \\
            \hline
        \end{tabular}
    \end{center}
\end{table}

\section{Conclusion}
In this study, we explored the ability of generative models for T1-FNC translation within the context of Alzheimer's disease. We employed cycle-GAN to synthesize FNC and T1 from each other, and to enhance the efficacy of our model, we integrated weak supervision during the transformation in cases where paired data were available. Our findings suggest that our proposed GAN-based approach can effectively capture and replicate intricate structural and functional brain patterns associated with Alzheimer’s disease. As an interesting future direction and application, one can investigate the potential of such a generative approach in downstream tasks such as diagnosis using multimodal data, by filing the missing modalities with the generated samples.

\addtolength{\textheight}{-12cm} 



\begin{thebibliography}{99}

\bibitem{vlaardingerbroek2013magnetic} M. T. Vlaardingerbroek and J. A. Boer, Magnetic Resonance Imaging: Theory and Practice. Springer Science \& Business Media, 2013.

\bibitem{logothetis2008we} N. K. Logothetis, “What we can do and what we cannot do with fMRI,” in Nature, vol. 453, no. 7197, pp. 869--878, 2008. Nature Publishing Group UK London.

\bibitem{phelps2000positron} M. E. Phelps, “Positron emission tomography provides molecular imaging of biological processes,” in Proceedings of the National Academy of Sciences, vol. 97, no. 16, pp. 9226--9233, 2000. National Acad Sciences.

\bibitem{calhoun2016multimodal} V. D. Calhoun and J. Sui, “Multimodal fusion of brain imaging data: A key to finding the missing link(s) in complex mental illness,” in Biological Psychiatry: Cognitive Neuroscience and Neuroimaging, vol. 1, no. 3, pp. 230--244, 2016. Elsevier.

\bibitem{venugopalan2021multimodal} J. Venugopalan, L. Tong, H. R. Hassanzadeh, and M. D. Wang, “Multimodal deep learning models for early detection of Alzheimer’s disease stage,” in Scientific Reports, vol. 11, no. 1, pp. 3254, 2021. Nature Publishing Group UK London.

\bibitem{cai2018deep} L. Cai, Z. Wang, H. Gao, D. Shen, and S. Ji, “Deep adversarial learning for multi-modality missing data completion,” in Proceedings of the 24th ACM SIGKDD International Conference on Knowledge Discovery \& Data Mining, 2018, pp. 1158--1166.

\bibitem{gao2021task} X. Gao, F. Shi, D. Shen, and M. Liu, “Task-induced pyramid and attention GAN for multimodal brain image imputation and classification in Alzheimer’s disease,” in IEEE Journal of Biomedical and Health Informatics, vol. 26, no. 1, pp. 36--43, 2021. IEEE.

\bibitem{cheng2021research} D. Cheng, N. Qiu, F. Zhao, Y. Mao, and C. Li, “Research on the modality transfer method of brain imaging based on generative adversarial network,” in Frontiers in Neuroscience, vol. 15, pp. 655019, 2021. Frontiers Media SA.

\bibitem{tiago2023domain} C. Tiago, S. R. Snare, J. Šprem, and K. McLeod, “A Domain Translation Framework With an Adversarial Denoising Diffusion Model to Generate Synthetic Datasets of Echocardiography Images,” in IEEE Access, vol. 11, pp. 17594--17602, 2023. IEEE.


\bibitem{dar2019image} S. U. H. Dar, M. Yurt, L. Karacan, A. Erdem, E. Erdem, and T. Cukur, “Image synthesis in multi-contrast MRI with conditional generative adversarial networks,” in IEEE Transactions on Medical Imaging, vol. 38, no. 10, pp. 2375--2388, 2019. IEEE.

\bibitem{yuan2020unified} W. Yuan, J. Wei, J. Wang, Q. Ma, and T. Tasdizen, “Unified generative adversarial networks for multimodal segmentation from unpaired 3D medical images,” in Medical Image Analysis, vol. 64, pp. 101731, 2020. Elsevier.


\bibitem{zhu2017unpaired} J. Zhu, T. Park, P. Isola, and A. A. Efros, “Unpaired image-to-image translation using cycle-consistent adversarial networks,” in Proceedings of the IEEE International Conference on Computer Vision, 2017, pp. 2223--2232.

\bibitem{wang2004image} Z. Wang, A. C. Bovik, H. R. Sheikh, and E. P. Simoncelli, “Image quality assessment: from error visibility to structural similarity,” in IEEE Transactions on Image Processing, vol. 13, no. 4, pp. 600--612, 2004. IEEE.

\bibitem{shrivastava2017learning} A. Shrivastava, T. Pfister, O. Tuzel, J. Susskind, W. Wang, and R. Webb, "Learning from simulated and unsupervised images through adversarial training," in Proceedings of the IEEE Conference on Computer Vision and Pattern Recognition, 2017, pp. 2107--2116.

\bibitem{jack2008alzheimer} C. R. Jack Jr, M. A. Bernstein, N. C. Fox, P. Thompson, G. Alexander, D. Harvey, B. Borowski, P. J. Britson, J. L. Whitwell, C. Ward, et al., “The Alzheimer’s disease neuroimaging initiative (ADNI): MRI methods,” in Journal of Magnetic Resonance Imaging: An Official Journal of the International Society for Magnetic Resonance in Medicine, vol. 27, no. 4, pp. 685--691, 2008. Wiley Online Library.

\bibitem{du2020neuromark} Y. Du, Z. Fu, J. Sui, S. Gao, Y. Xing, D. Lin, M. Salman, A. Abrol, M. A. Rahaman, J. Chen, et al., “NeuroMark: An automated and adaptive ICA based pipeline to identify reproducible fMRI markers of brain disorders,” in NeuroImage: Clinical, vol. 28, pp. 102375, 2020. Elsevier.

\bibitem{goodfellow2020generative} I. Goodfellow, J. Pouget-Abadie, M. Mirza, B. Xu, D. Warde-Farley, S. Ozair, A. Courville, and Y. Bengio, "Generative adversarial networks," in Communications of the ACM, vol. 63, no. 11, pp. 139--144, 2020. ACM New York, NY, USA.

\bibitem{armanious2019unsupervised} K. Armanious, C. Jiang, S. Abdulatif, T. Küstner, S. Gatidis, and B. Yang, “Unsupervised medical image translation using cycle-MedGAN,” in 2019 27th European Signal Processing Conference (EUSIPCO), 2019, pp. 1--5. IEEE.

\bibitem{penalba2023increased} L. Penalba-Sánchez, P. Oliveira-Silva, A. L. Sumich, and I. Cifre, "Increased functional connectivity patterns in mild Alzheimer’s disease: A rsfMRI study," in Frontiers in Aging Neuroscience, vol. 14, pp. 1037347, 2023. Frontiers.

\bibitem{badhwar2017resting} A. Badhwar, A. Tam, C. Dansereau, P. Orban, F. Hoffstaedter, and P. Bellec, "Resting-state network dysfunction in Alzheimer's disease: a systematic review and meta-analysis," in Alzheimer's \& Dementia: Diagnosis, Assessment \& Disease Monitoring, vol. 8, pp. 73--85, 2017. Elsevier.

\bibitem{mckhann2011diagnosis} G. M. McKhann, D. S. Knopman, H. Chertkow, B. T. Hyman, C. R. Jack Jr, C. H. Kawas, W. E. Klunk, W. J. Koroshetz, J. J. Manly, R. Mayeux, et al., "The diagnosis of dementia due to Alzheimer’s disease: Recommendations from the National Institute on Aging-Alzheimer’s Association workgroups on diagnostic guidelines for Alzheimer's disease," in Alzheimer's \& Dementia, vol. 7, no. 3, pp. 263--269, 2011. Elsevier.


\end{thebibliography}
\end{document}